\documentclass[letterpaper, 10 pt, conference]{ieeeconf}

\makeatletter
\newcommand*\titleheader[1]{\gdef\@titleheader{#1}}
\AtBeginDocument{%
  \let\st@red@title\@title
  \def\@title{%
  	\vskip-2.4em
    \bgroup\normalfont\large\centering\@titleheader\par\egroup 
    \vskip1.4em\st@red@title}
}
\makeatother  

\IEEEoverridecommandlockouts                              
\usepackage{graphics} 
\usepackage{epsfig} 	
\usepackage{mathptmx} 	
\usepackage{times} 		
\usepackage{amsmath} 	
\usepackage{amssymb}  	
\usepackage{tabularx}
\usepackage{color}
\usepackage[utf8]{inputenc}
\usepackage[bookmarks=false]{hyperref}

\title{\LARGE \bf
Temporal orders and causal vector for physiological data analysis
}

\titleheader{\color{red} \small \textit{Draft accepted for the \textbf{42nd IEEE Annual International Conference of the IEEE Engineering in\\ Medicine and Biology Society, EMBC'20}. DO NOT REDISTRIBUTE.}}

\author{Marcel Młyńczak$^{1}$%
\thanks{$^{1}$M. Młyńczak is with the Warsaw University of Technology, Faculty of Mechatronics, Institute of Metrology and Biomedical Engineering, 8 Boboli Street, 02-525, Warsaw, Poland. {\tt\small marcel.mlynczak@pw.edu.pl}}%
}

\begin{document}

\maketitle
\thispagestyle{empty}
\pagestyle{empty}

\begin{abstract}
In addition to the global parameter- and time-series-based approaches, physiological analyses should constitute a local temporal one, particularly when analyzing data within protocol segments. Hence, we introduce the R~package implementing the estimation of temporal orders with a causal vector (CV). It may use linear modeling or time series distance. The algorithm was tested on cardiorespiratory data comprising tidal volume and tachogram curves, obtained from elite athletes (supine and standing, in static conditions) and a~control group (different rates and depths of breathing, while supine). We checked the relation between CV and body position or breathing style. The rate of breathing had a greater impact on the CV than does the depth. The tachogram curve preceded the tidal volume relatively more when breathing was slower.
\newline

\indent \textit{Clinical relevance}— The method can assess (1) relationships between two signals, having one of the two time-shifted in a~given range, (2)  curves of most optimal inter-signal shift, and (c) their stability across time; also for other physiological studies.
\end{abstract}

\section{INTRODUCTION}

The network physiology concept rests on the assumption that several systems are combined in a network of dependencies with various loops, feedbacks, and delays in transmitting information (also influenced by many environmental, psychological, or demographic factors) \cite{Bartsch2015,Fatisson2016}.  

One of the combinations is that of respiration and heart activity. It appears multi-directional and very intricate. Respiratory sinus arrhythmia (RSA) is a phenomenon evident in heart rate modulation as the effect of successive inspirations and expirations \cite{Larsen2010}. On the other hand, cardiorespiratory coupling holds that cardiac activity triggers respiration, particularly because of increased sympathetic nervous activity and afferent baroreflex discharge \cite{Larsen2010,Reyes2013,Penzel2016}.  

Many mathematical approaches have been proposed to analyze relationships. We hypothesized (primarily from the sports medicine, normative perspective) that the parameterization of the cause-and-effect relationships would also be valuable. Therefore, we assumed that the analysis should comprise (1) global parameters examination without considering the impact of time; then (2) it should delve into global temporal (time series) relationships and causalities; and finally, (3) local analysis of narrow time intervals would be performed for the directionality, strength and stability assessments within each protocol segments (e.g., static supine as the first step of orthostatic maneuver).

The first two approaches (global time-independent, and global time-dependent) have already been checked in \cite{Mlynczak2018,Mlynczak2019}. In the former one, we tried to discover causal paths, and the results suggested body position is an essential factor to be taken into account. To bring a context,  in a supine position, the values of a tidal volume seemed to cause heart activity variation, which affected average heart activity, and the chain was completed with respiratory timing parameter. For standing, in turn, the relation led from normalized respiratory activity variation to average heart activity \cite{Mlynczak2018}. 

In the latter approach, temporal relationships were examined by (a) Granger causality frameworks (with extensions that consider zero-lag effects \cite{Schiatti2015}) and (b) Time Series using Restricted Structural Equation Models (TiMINo) \cite{Peters2013}. The most prominent "global" combination appeared between the tachogram (RR intervals curve) and tidal volume signal \cite{Mlynczak2019}, as a result of RSA. However, the results were weak and unstable, which is probably an effect of considering relatively long segments of data. But, on the other hand, temporal causality analyses require minimum lengths of data to work correctly, e.g., by definition and because of the stationarity criterion. 

Therefore, there is a~need for a~framework to examine local, short segments of data and to explore the temporal orders (TO) between signals. The introduction and preliminary evaluation of such a technique is hence the main aim of this study.

\section{MATERIALS AND METHODS}

\subsection{Description and interpretation of the algorithm}

The method is implemented as an R package, "tempord", supplementing the paper. The "installer", the CRAN-compatible manual, and the scheme of the algorithm are available on the author's website \cite{MM}. The package uses external R packages: \textit{ggplot2} and \textit{TSdist}.

The bivariate dataset should be loaded. The first signal is kept stationary during the analysis, while another is shifted in time (backward and forward, in the given range). 

The most important parameters that define the entire setup (all are described in the R package manual), are: 

\begin{itemize}
	\item method - determines which approach will be used (linear modeling or time series distance; further explained below);
	\item scaling - states the standardization type: 0 - no standardization, 1 - uniform standardization, or 2 - Gaussian standardization; 
	\item length of a signal segment to analyze per loop iteration (the same length for both signals; should not exceed the length of the entire signal); and
	\item maximum shift in time - determines how far the second signal may be moved backward and forward; the range does not need to be symmetrical to zero shift.
\end{itemize}

The main loop of the analysis is carried out for each point of the first signal with a shift applied to the second signal (according to the set resolutions). In the first step, the selection of the signal parts is taken for analysis during each iteration. Then follows scaling if selected.

The main parameters are: (1) the adjusted R-Squared measure, for linear modeling (LM) approach, or (2) the time series distance (TD) measure (e.g., Manhattan or Fourier types), for another, TD approach.

After all iterations, the matrix is filled and may be presented in the temporal orders graph. To facilitate assessing stabilities, the curve of maximum or minimum values (for all shifts considered) for each point in time (for linear modeling and time series distance, respectively) - hereafter called causal vector (CV), is calculated. If a~threshold (the level at which the estimate of the parameter is excluded when determining CV) is chosen, only the values (respectively) above or below the threshold will produce those curves.

The setting of the right shift range is crucial. If the range is too wide, the impact of consecutive periods would be visible. Also, in general, the causal vector curve should be non-intermittent (when considering thresholds) and near-parallel to the X-axis. In such fragments of the signal, the temporal relation may be regarded as stable. Step changes in the course of the maximum values may reflect extrinsic changes in the study conditions.

\subsection{Study groups and the protocol}

The participants created two groups (demographically described in Table I):

\begin{itemize}
	\item 10 elite athletes (A) - 4 minutes of registrations for supine and standing body positions, with unconstrained and free breathing protocols (sample choice from the study group described in  \cite{Mlynczak2018,Mlynczak2019});
	\item a control group of 10 students (B) - following a~constrained breathing procedure comprising 12 breaths (6~normal, then 6 deep) each at rates of 6, 10, and 15 breaths per minute (BPM).
\end{itemize}

The experimental procedures involving human subjects described in this paper were performed according to the principles outlined in the Helsinki Declaration and the part regarding the athletes also approved by the Ethics Committee of Warsaw Medical University (permission AKBE/74/17).

\begin{table}[!h]
\centering
\caption{The demographic summary of both study groups.}
\begin{tabularx}{\columnwidth}{X|rr}
\hline
\textbf{Group} & \textbf{A} & \textbf{B}  \\ \hline
\textbf{Count} & $10$ & $10$  \\
\textbf{Sex} & 5F \& 5M & 3F \& 7M  \\
\textbf{Weight [kg]} & $68.5 \pm 15.3$ & $72.9 \pm 13.2$  \\
\textbf{Height [cm]} & $179.5 \pm 12.6$ & $176.2 \pm 7.2$  \\ \hline
\end{tabularx}
\end{table}

Single-lead ECG and impedance pneumography signals were acquired using the Pneumonitor 2 \cite{Pneumonitor}. The tachogram was calculated from the ECG signal using the identification of R~peaks based on the Pan-Tompkins algorithm. A tidal-volume-related impedance signal was obtained without calibration (only relative shape matters, as linear fitting, provides the best agreement between the impedance signal and the reference, pneumotachometry \cite{Mlynczak2015}). We reduced the sampling frequency from the original $250Hz$ to $25Hz$ for computational reasons.

\subsection{Cardiorespiratory analysis}

The algorithm was applied to all the signals to find the graphs of temporal orders with the courses of a causal vector. Then, the CVs were analyzed concerning body positions and changes in breathing style, to determine how they affect the relationships.

During the analysis, average and standard deviation values of the estimated causal vector were calculated for supine/standing body positions in Group A, and for different depths and rates of breathing in Group B, using the main settings: 

\begin{itemize}
	\item with Gaussian standardization;
	\item segment length - 10 seconds; and
	\item maximal shift in both directions - 2 seconds (based on average breathing frequency and the RSA phenomenon characteristics).
\end{itemize}

Then, to assess the stability more robustly, we checked the duration of the longest stable parts and the duration of CV to that of the entire signal. If we apply thresholds, $0.90$ for LM and $0.15$ for TD approaches will be used arbitrarily.

\section{RESULTS}

Sample temporal order graphs for very regular signals recorded for a supine position are shown in Fig. 1, and for less regular signals obtained during standing - in Fig. 2. In both cases, we applied thresholds. 

\begin{figure}[!h]
\centerline{\includegraphics[width=\columnwidth]{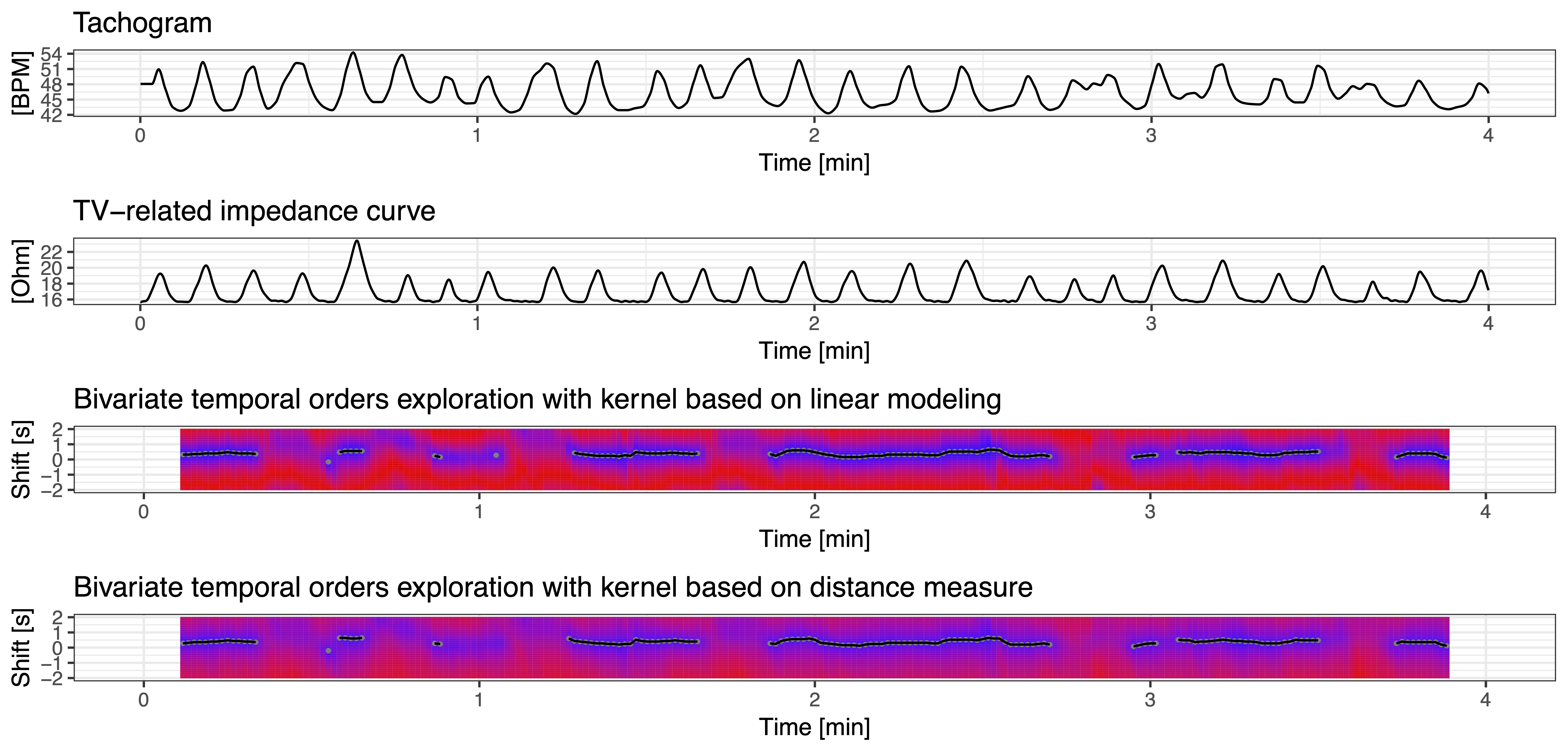}}
\caption{Sample TO estimation for very regular signals acquired from Group A participant \#1 (supine). For LM: navy blue means 1 (best) and red means 0 (worst); for TD: navy blue means the smallest possible value (best).}
\end{figure}

\begin{figure}[!h]
\centerline{\includegraphics[width=\columnwidth]{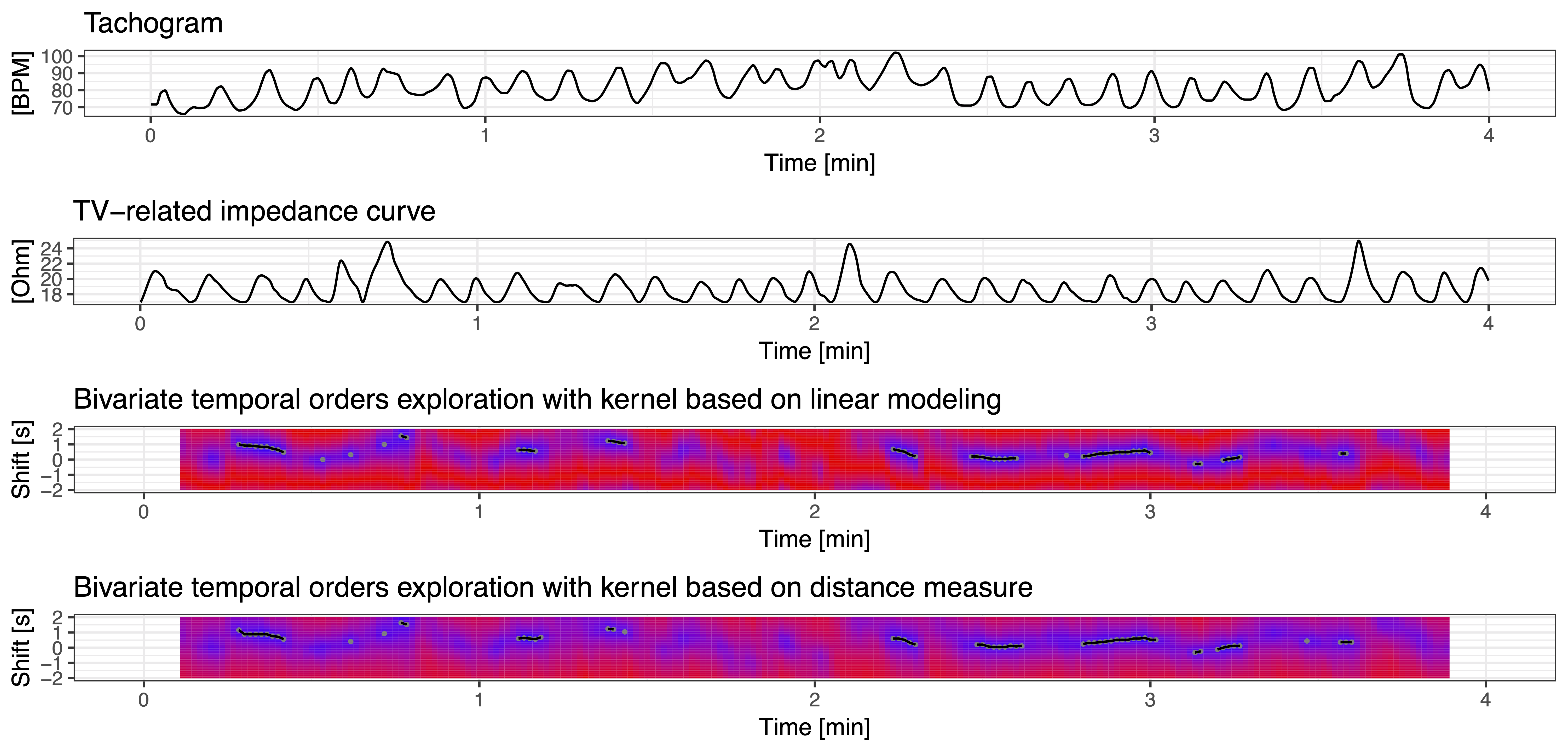}}
\caption{Sample TO estimation for less regular signals acquired from another Group A participant \#3 (standing). Colorbars description is the same as for Fig. 1.}
\end{figure}

Table II gathers the means and standard deviations of the causal vector in Group A for both supine and standing body positions without a threshold, for linear modeling, and TD Fourier kernels. The effect of respiratory sinus arrhythmia seems to be present, particularly for the supine body position. Alternative results obtained using considered methods come from different definitions, which may affect their compliance.

\begin{table}[!h]
\centering
\caption{Summary of the mean $\pm$ SD of causal vector value depending on body position, for all participants in Group A. A minus means that the tidal volume curve is ahead of the tachogram. All values are in milliseconds.}
\begin{tabularx}{\columnwidth}{X|rr}
\hline
\textbf{Position} & \textbf{Linear modeling} & \textbf{TD Fourier-type} \\ \hline
Supine & $302 \pm 466$ & $469 \pm 518$  \\
Standing & $-2 \pm 244$ & $239 \pm 794$  \\ \hline
\end{tabularx}
\end{table} 

Table III collects the duration of the longest stable CV part and ratio of the duration to that of the signal) after applying a threshold only to the TD-Manhattan approach.

\begin{table}[!h]
\centering
\caption{Summary of the mean $\pm$ standard deviations of the duration of the longest stable CV part, and ratio of the duration to the duration of the signal, for all participants in Group A, depending on body position. "N" counts, in how many cases the value above the threshold was found.}
\begin{tabularx}{\columnwidth}{X|l|rr}
\hline
\textbf{Position} & \textbf{N} & \textbf{Longest [sec]} & \textbf{Ratio [\%]} \\ \hline
Supine & $7$ & $15.1 \pm 16.8$ & $21.2 \pm 20.1$  \\
Standing & $6$ & $9.0 \pm 4.0$ & $11.0 \pm 8.8$  \\ \hline
\end{tabularx}
\end{table} 

When applying arbitrary thresholds, there are relatively few and short fragments of constant causal vectors between signals. The mean separate causal vectors lasted $571ms$ and $431ms$ for supine and standing body positions, respectively.

Next, sample visual results for the entire breathing protocol in Group B are shown in Fig. 3. 

\begin{figure}[!h]
\centerline{\includegraphics[width=\columnwidth]{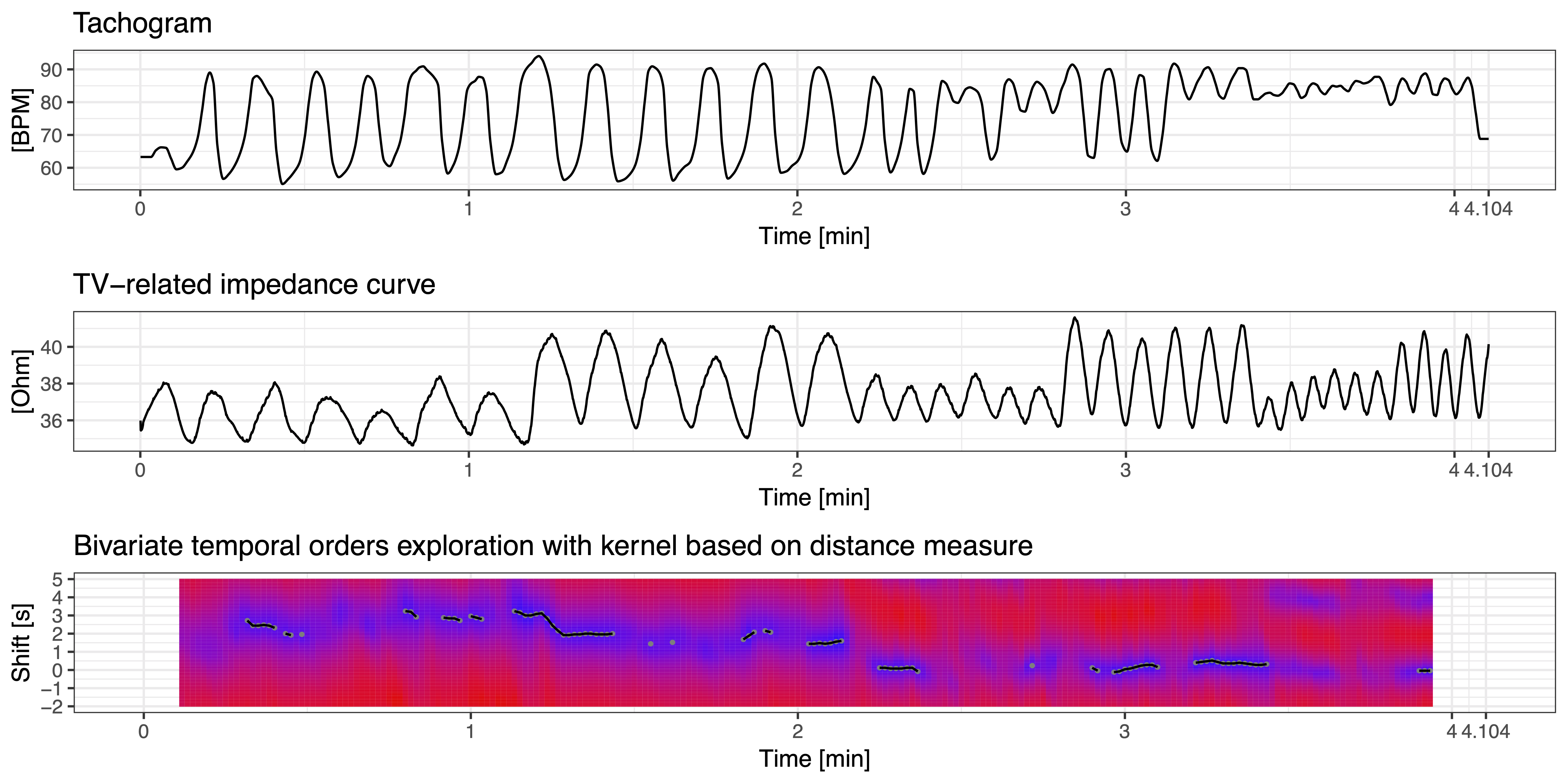}}
\caption{Sample TO estimation for the whole protocol acquired from a~Group B participant \#7; using only the TD-Manhattan approach with a~threshold, with shifts from $-2s$ to $5s$. For the last part of the signal (15 BPM breathing rate), the second period appears to be present for shifts close to $5s$. Colorbars description is the same as for Fig. 1.}
\end{figure}

Similarly, as earlier, the mean and standard deviations were calculated for the TD-Manhattan approach (without thresholds) for different breathing rates and depths (in Group B) and are shown in Table IV. 

\begin{table}[!h]
\centering
\caption{Summary of mean $\pm$ standard deviations of causal vector values, for all Group B, for every combination of breathing depth (normal and deep) and breathing rate (6, 10, and 15 breaths per minute, BPM), for the TD Manhattan approach, without a threshold. All results are in milliseconds.}
\begin{tabularx}{\columnwidth}{X|rrr}
\hline
\textbf{Depth} & \textbf{6 BPM} & \textbf{10 BPM} & \textbf{15 BPM} \\ \hline
Normal & $2469 \pm 643$ & $829 \pm 280$ & $323 \pm 334$  \\
Deep & $2499 \pm 380$ & $854 \pm 363$ & $304 \pm 317$  \\ \hline
\end{tabularx}
\end{table} 

The results suggest the rate of breathing has a more significant impact on the mean value of the causal vector than the depth of breathing. The comparison of the average CVs (calculated as a mean value independently of the depth) for Group B, along with the cycle durations, depending on the breathing rate, are stored in Table V. 

\begin{table}[!h]
\centering
\caption{The comparison of average causal vector values, calculated independently of the breathing depth for all Group B, with the cycle duration and the ratio between optimal shift (average causal vector), depending on the breathing rate (6, 10, and 15 breaths per minute, BPM), for the TD-Manhattan approach, without a threshold.}
\begin{tabularx}{\columnwidth}{X|rrr}
\hline
\textbf{Rate} & \textbf{Optimal shift (OS)} & \textbf{Cycle duration (CD)} & \textbf{OS/CD} \\ \hline
6 BPM & $2484ms$ & $10s$ & $24.8\%$  \\
10 BPM & $842ms$ & $6s$ & $14.0\%$  \\ 
15 BPM & $314ms$ & $4s$ & $7.9\%$  \\\hline
\end{tabularx}
\end{table} 

Interestingly, the RR interval curve precedes the tidal volume more when breathing is slower. It is further discussed in the next section.

\section{DISCUSSION}

The causal analysis appears to be a promising tool to extend the classical approaches of cardiorespiratory analysis and enable answers to new physiological questions. In a~traditional approach (if considered at all), a graph of connections is rather taken as an input, "prior information" based on medical knowledge, even if it is relatively hard to be established, specifically for an individual athlete. We would like to emphasize the opposite approach, in which the results of the causal-related analysis might be used as the input. We call this the "bottom-up" strategy \cite{Mlynczak2018,Mlynczak2019}.

It can be used by practitioners to (1) identify optimal training schedules to promote desirable progress and achieve optimal performance (based on the individual directions and strengths of cardiorespiratory relationships), or to (b) establish sufficient training loads; all considering interventions applied to the cause "variables" \cite{Mlynczak2018,Mlynczak2019}.

Moreover, the study protocols might be arranged in a~way enabling testing of adaptation, recovery status, etc. Then, the possible levels of analysis might be supplemented with a local temporal order estimation. In our opinion, the presented concept may be used primarily to test the short temporal stability of relationships (this has already been discussed by Porta et al. \cite{Porta2018}, in the context of different physiological phenomena), but also to check their directionality and strength. The stability can be described as strong when the causal vector is almost constant over time. Protocol changes, like switching from supine to standing, may be compared regarding explored temporal orders. Directionality, in turn, can be indirectly assessed by the signal shift sign.

The results were that RR interval curve precedes the tidal volume one. It is not contradictory to RSA but it results from taking signals into consideration. The local maximum of RR interval occurs during the inspiratory phase, so based on the shape, it is earlier, still being caused by respiratory activity. Worth noted is that if tidal volume was replaced by airflow signal as derivative, temporal orders would change.  

We then also observed (in the control group, where controlled breathing removes effects other than RSA) that the precedence of the tachogram curve before the tidal volume one is relatively greater when the breathing is forced to be slower. However, as the presented analysis is only an illustration, we do not treat it as a general medical conclusion. This issue certainly requires a further look. In our opinion, this may be connected to the slow breathing concept as an intervention that subjects may practice benefiting health (particularly cardiovascular) \cite{Russo2017}. Here, the relation between the depth of breathing and causal vector values appeared insignificant; however, to maintain a~lower breathing rate without disturbing respiratory homeostasis, a tidal volume must be increased (the relation between temporal and amplitude ventilation coefficients \cite{Russo2017}).   

We think various improvements may be considered:

\begin{itemize}
	\item introducing weights for different depths of shifts;
	\item assessing temporal orders not only based on raw data, but also on signals' phases;
	\item analyzing the series of sub-maximum values during a~stable segment of data to indirectly evaluate the local strength of connections (not only a causal vector);
	\item considering more sophisticated approaches;
	\item transitioning an estimation to the multivariate case; 
	\item evaluating not only the signals' shapes but also their local extrema (enabling more to find phase relations and weaker and intermittent effects); or
	\item enabling a more robust (perhaps even automatic) choice of the method's parameters, e.g., the thresholds for both LM and TD.
\end{itemize} 

The method could be a useful complement, e.g., for transfer entropy (used in bivariate case); however, the limitations of the presented technique are (1) the relatively substantial dependence on the input parameter settings (in the presented setup they were manually fixed), and (2) the significance of the choice of analyzed signals. They should be carefully adapted or considered based on the problem.

\section{CONCLUSIONS}

A technique for exploring temporal orders in physiological data was introduced using the example of cardiorespiratory data (tachogram and tidal volume curves) in two groups (elite athletes and control, made up of students). We proposed two different approaches. The former uses linear modeling, the latter time series distance calculation. 

Respiratory sinus arrhythmia effect seems to be the most visible, particularly for the supine body position. The results also suggested that the rate of breathing has a greater impact on the mean value of the causal vector than does the depth of breathing. Interestingly, the RR interval curve precedes the tidal volume more when breathing is slower.  

The method can be used as an initial step during causality analysis, as it may show the temporal relationships between signals, the stability of the temporal orders over time, highlighting the similar parts of the causal vector curve (which can be then compared with events in the study protocol).


\end{document}